\documentstyle[eqsecnum,pre,preprint,aps,epsfig]{revtex}

\tightenlines

\newcommand{\bq}{\begin{equation}}
\newcommand{\eq}{\end{equation}}
\newcommand{\bqa}{\begin{eqnarray}}
\newcommand{\eqa}{\end{eqnarray}}
\newcommand{\ben}{\begin{enumerate}}
\newcommand{\een}{\end{enumerate}}
\newcommand{\bc}{\begin{center}}
\newcommand{\ec}{\end{center}}
\newcommand{\bqb}{\begin{eqnarray*}}
\newcommand{\eqb}{\end{eqnarray*}}
\newcommand{\psl}{p\hskip-0.21cm\slash}

\begin{document}

\draft
\preprint{PM/02-54}

\title{\vspace{1cm}  Charged Higgs Production in the 1 TeV Domain as a \\
Probe of Supersymmetric Models
\footnote{Partially supported by EU contract HPRN-CT-2000-00149}}
\author{M. Beccaria$^{a,b}$,
F.M. Renard$^c$, \\
S. Trimarchi$^{d,e}$ and C. Verzegnassi$^{d, e}$ \\
\vspace{0.4cm}
}

\address{
$^a$Dipartimento di Fisica, Universit\`a di
Lecce \\
Via Arnesano, 73100 Lecce, Italy.\\
\vspace{0.2cm}
$^b$INFN, Sezione di Lecce\\
\vspace{0.2cm}
$^c$ Physique
Math\'{e}matique et Th\'{e}orique, UMR 5825\\
Universit\'{e} Montpellier
II,  F-34095 Montpellier Cedex 5.\hspace{2.2cm}\\
\vspace{0.2cm}
$^d$
Dipartimento di Fisica Teorica, Universit\`a di Trieste, \\
Strada Costiera
 14, Miramare (Trieste) \\
\vspace{0.2cm}
$^e$ INFN, Sezione di Trieste\\
}

\maketitle

\begin{abstract}
We consider the production, at future lepton colliders,
of charged Higgs pairs in supersymmetric models. Assuming
a relatively light SUSY scenario, and working in the MSSM,
we show that, for c.m. energies in the one TeV range, a one-loop
logarithmic Sudakov expansion that includes an "effective" 
next-to subleading order term is adequate to the expected
level of experimental accuracy. We consider then the coefficient 
of the linear (subleading) SUSY Sudakov logarithm and the SUSY 
next to subleading term of the expansion and show that their 
dependence on the supersymmetric parameters of the model
is drastically different. In particular the coefficient of the SUSY
logarithm is only dependent on $\tan\beta$ while the next to
subleading term depends on a larger set of SUSY parameters. This
would allow to extract from the data separate informations 
and tests of the model.

\end{abstract}

\section{Introduction}

In the last few years, a considerable effort has been devoted 
to the precise formulation             
of the theoretical predictions for electroweak effects in pair 
production at future lepton colliders. In particular, the considered 
center of mass (c.m.) energies have been those
that represent the final goal of two proposed future machines, 
roughly one TeV for LC \cite{LC} and three TeV for CLIC \cite{CLIC}. 
The main motivation of the various investigations has been the fact
that, within the electroweak sector of the SM, for c.m. energies 
of the few TeV size, it has been
realized \cite{large,log,susylog} that unexpectedly   
large virtual effects arise at the one loop level, 
that could make the validity of this (relatively) 
simple perturbative calculation highly debatable. 
These terms have the analogous dependence
on energy as those originally determined in QED by Sudakov
\cite{Sudakov}; at one loop, they can either be of
squared logarithmic (leading) (DL) or of linear logarithmic
(subleading) (SL) kind, 
and their numerical effect in
several observables breaks the "safety" few percent limit (fixed
by the aimed (~one percent) experimental accuracy) when one enters the 
few (2-3) TeV region, making the request of a higher order 
calculation to become imperative in that range.\\

Within the SM framework a resummation to all orders actually 
exists to \underline{next-to subleading order}
for final massless fermion pairs, 
and to \underline{subleading order} for general massive 
final pairs~\cite{resum1}. In the (particularly relevant) case of 
\underline{massive} final fermions a comparison 
(to subleading logarithmic accuracy) between one-loop 
and resummed expansions has been performed~\cite{resum2}.
The results indicate that, for c.m. energies entering the few 
(2-3) TeV range, the discrepancies between the
two approximations become intolerably (i.e. beyond a relative
ten percent) large. On the contrary, for energies in 
the one TeV region, no 
appreciable difference shows up: to subleading order, 
the one-loop description appears there adequate. 
The question remains that of whether possible next-to subleading 
(for instance constant) terms might play a role.\\

For massive (third generation) final quark pairs, this problem 
was investigated in an 
``effective'' way~\cite{top}, trying to fit the exact one-loop 
calculation
of a simple class of diagrams with a logarithmic expansion also 
containing an additional
constant. The result was that, in the one TeV region, this fit 
was adequately describing the exact calculation, with the 
logarithmic coefficients exactly predicted by the 
Sudakov expansion and a constant term given by the fit. 
The size of this term was (relatively) ``large'', 
about one half of the logarithmic contribution, with opposite sign
(thus decreasing the overall effect). The apparent 
conclusion was that, in that energy range,
a one-loop Sudakov expansion \underline{implemented 
by an extra constant term} seems to be
able to reproduce the exact calculation with an accuracy 
that is largely sufficient, at the expected experimental 
precision level of a relative one percent.

As a comment on the previous conclusion, it can be noticed 
that the possible experimental determination
of the separate logarithmic and constant terms would not 
lead to any new information in the SM, since all the various 
coefficients will depend on the (fixed) known values of the 
SM parameters. In this sense, accurate measurements of 
massive (and massless) fermion pair production at LC in 
the one TeV region can only provide, in the SM framework, 
another experimental test of the validity of the model.

A natural question that arises at this stage is that of 
whether similar, or different, conclusions
can be drawn for the simplest supersymmetric generalization 
of the SM that can still be treated perturbatively, i.e.
the Minimal Supersymmetric Standard Model. This model and 
its Sudakov expansion has been actually already considered 
to subleading logarithmic accuracy in a number of papers, to the
one-loop level for massless and massive fermion
pairs~\cite{log,susylog} 
and resumming to all orders for sfermion and Higgs 
production~\cite{scalar}. In the latter case, 
a comparison between the one-loop and the resummed expansion 
has also been performed under the assumption of a 
relatively light SUSY scenario, showing that, in strict analogy  
with the SM situation, the two calculations
are essentially identical in the one TeV region,
while deep differences show up in the (2,3) TeV energy range. 
No effort was made to try to estimate
the size of a possible next to subleading term in the one-loop 
expansion in the one TeV region from a fit to the exact 
calculation, analogous to that performed in ~\cite{top} 
for the SM case. \\

The aim of this paper is precisely that of studying the 
feasibility and the possible advantages of performing an 
effective logarithmic one-loop expansion, implemented
by a next-to subleading term, in the energy region around 1 TeV, 
for the MSSM. We anticipate that a preliminary necessary 
condition will be that of a light SUSY scenario, in which all the 
SUSY masses relevant for the considered process are supposed 
to be not heavier than a few hundred GeV. This means that the starting 
picture will be one where SUSY has already been discovered 
via direct production of (at least some) sparticles, 
so that a number of SUSY masses is already known and fixed.
This is not obviously true for other quantities of the model, 
like for instance $\tan\beta$ or other parameters of the Higgs 
sector.\\

The purpose of our investigation will be that of showing that 
one fundamental differences will arise
in the MSSM analysis with respect to the SM case. 
Our conclusion will be in fact that the coefficient of the SUSY
Sudakov logarithm and the next to subleading term will exhibit a
drastically different dependence on the parameter of the model, 
so that their possible experimental identification might lead 
to quite valuable information. In particular, 
we shall reconsider, in this new spirit, the possibility 
of a determination of $\tan\beta$, that is the only
SUSY parameter on which the coefficient of the SUSY 
Sudakov (linear) logarithm depends,
as already shown in previous papers~\cite{scalar,tgbeta}. 
Also, we shall 
show that it is possible to investigate separately the dependence on
the next to subleading term on the remaining SUSY parameters 
in the light scenario that we have assumed. In fact, we shall also
show in some detail what is the range of SUSY parameters that may be
considered "light", from the point of view of a Sudakov expansion
like ours, in the one TeV region.\\

We should anticipate at this point the reason why we insist in calling
the remaining, non logarithmic component of the asymptotic one-loop
expansion "next to subleading" term. In fact, we shall show that this
remaining component cannot be \underline{rigorously} considered as a
constant in the investigated TeV region. To reproduce completely the
exact calculation, one must add to the constant component an extra non
logarithmic energy dependent quantity. As we shall show, though, this
extra component is, at the expected level of accuracy, fairly "small".
The consequence of its presence will be fully taken into account 
and will generate a realistic error in the determination of the SUSY 
parameter that we shall pursue.\\

As a first process to be examined in this spirit, we have chosen 
that of charged Higgs pair production.
The main reason for this choice is that, from the point of view 
of the involved parameters, this is the simplest process to be 
considered in the MSSM. Our goal is that of moving 
in future papers to more complicated processes, following a 
logical chain that introduces gradually new parameters
not already derived in this effective way. In order to provide a
self-consistent and rigorous calculation device, we have completed a
one loop code that contains all the relevant diagrams, in the
approximation of treating all fermions as massless, with the
exception of the third generation quarks. We have verified that the
complete code does reproduce the correct(known) asymptotic
logarithmic (DL and SL) behaviour. This complete code has been
used to derive an effective next-to subleading Sudakov expansion, with
which it has been imposed to it to agree to the few permille level.
The code, that has been named SESAMO (Supersymmetric Effective
Sudakov Asymptotic MOde), is already available for use~\cite{SESAMO}.\\\\

Technically speaking, this paper will be organized as follows: 
Sec. II will contain the relevant asymptotic one loop 
expansions  of the process. Sec. III will contain the comparison
of the complete one loop calculation with the proposed effective fit
as a function of the SUSY parameters. In Sec. IV, 
the determination of $\tan\beta$ and the study of the effect of
the remaining parameters on the next-to
subleading term will be examined and exhibited. A final discussion
of our results in Sec. V will conclude the paper.

\section{Calculation of the process at one loop}

A complete description of the scattering amplitude of the considered
process at one loop requires the calculation of several classes of
diagrams. To make our treatment as self-consistent as possible, we
shall follow the notations used in a previous
reference~\cite{scalar} and write:

\bqa
A(e^+e^-\to H^+H^-)&=&A^{{\rm Born}}(e^+e^-\to H^+H^-)+
A^{c.t.}(e^+e^-\to H^+H^-)\nonumber\\
&&+A^{s.e.}(e^+e^-\to H^+H^-)
+A^{in}(e^+e^-\to H^+H^-)\nonumber\\
&&
+A^{fin}(e^+e^-\to H^+H^-)+A^{box}(e^+e^-\to H^+H^-)
+A^{QED}
\label{A}\eqa

It is convenient to normalize the general amplitude in
the following way:

\bq
A={2e^2\over q^2}~\bar v(e^+)(\psl)(a_LP_L+a_RP_R)u(e^-)
\label{aLR}\eq
\noindent
where $P_{L,R}=(1\mp\gamma^5)/2$ and $p^{\mu}$ is the outgoing
$H^-$ momentum, so that one writes the Born terms as:

\bq
a^{{\rm Born}}_{\lambda}=1-{(1-2s^2_W)\over4s^2_Wc^2_W}\eta g_{e\lambda}
\eq
\noindent
with
$g_{eL}=2s^2_W-1,~~g_{eR}=2s^2_W$ 
and $\eta\equiv {q^2\over q^2-M^2_Z}$.

The remaining quantities represent the one-loop perturbative
modifications of the tree level expression. More precisely, $A^{c.t.}$
is the contribution from the usual "counter-terms" that cancel all the
ultraviolet divergences of the process. In our chosen
on-shell renormalization scheme, in which the inputs are
$\alpha_{QED}(0)$, $M_W$ and $M_Z$, 
they are given by proper gauge boson
self-energies, computed at the corresponding physical masses. Their
explicit expressions are known (see e.g. \cite{Hollik}), and to save
space we shall note write them here. $A^{s.e.}$ contains the various
internal self-energy corrections; $A^{in,~fin}$ describe the initial 
and final vertex modifications and $A^{box}$ the box contributions. 
Their "fine" structure is summarized in Appendix, where the various 
components of the separate terms are listed. The related complete 
set of Feynman diagrams is too large (more than 200 diagrams) to be
drawn here; it can be found e.g. in a recent paper \cite{HHHol}. 
In our treatment, we have discarded those components
that give contributions that vanish with the initial lepton mass,
which reduces somehow our numerical calculations; (actually, we
treated all fermions as massless, with the exception of the third
family quarks). In our approach, ultraviolet divergences are produced
both by self-energies and by vertices that include all
\underline{external} self-energies (we followed the definition proposed
by Sirlin \cite{DS}). We checked that all the ultraviolet
divergences in the scattering amplitude are mutually canceling.
Finally, $A^{QED}$ represents our choice of the electromagnetic
component. In this preliminary paper, we were mostly interested in the
"genuine" electroweak SUSY contributions at very high energies. For
this reason, we treated all those virtual contributions with photons
that would generate infrared divergences by introducing an "effective"
fictitious photon mass $M_{\gamma}=M_Z$. With this choice, our
$A^{QED}$ will contain in fact the difference between these
"effective" terms and the conventional (massless photon) ones (with
the usual request of adding the effects of real photon radiation when
computing the cross section). This gives a contribution to the
observables that, for our specific purposes, can be considered as
"known" since it does not involve any SUSY component, and will
therefore not be included in our code at least at the moment. Its
addition would not represent a problem, but would not add anything to
the specific investigation of this paper, that is rather devoted to
the "unknown" component of the scattering amplitude.\par
In this spirit, we shall now write the \underline{overall} logarithmic
contributions to the scattering amplitude that arise at one loop in a
proper configuration of "asymptotic" energy. These are by definition
the leading terms of an expansion made in a region where
\underline{all} the relevant (external and internal) masses are
sufficiently smaller than the c.m.  energy in a way that we shall 
try to make more quantitative in the following Section. In full 
generality, such logarithmic terms can be of two different origins. 
The first ones are linear logarithms of renormalization group (RG) 
origin, generated by gauge boson self-energies and representing 
the "running" of the gauge coupling constants. Their expressions 
are known, and we shall write them explicitely but separately from the
remaining terms. They do not contain any SUSY parameter, but must be
carefully taken into account in our approach. The second ones are the
genuine electroweak logarithms,  nowadays generally called 
"of Sudakov type". They can be of quadratic and of linear kind; the
quadratic ones come from vertices with single $W,~Z,~\gamma$ exchange
and from boxes with two $W$ exchange and \underline{do not}
contain any SUSY parameter; the linear ones come from the remaining
vertices and boxes and contain SUSY contributions only from vertices.
A very special feature of the supersymmetric model that we have
investigated and of the process that we have considered is that the
only SUSY parameter that "effectively" appears in the various linear
logarithms is $\tan\beta$, of Yukawa origin, produced by
the final vertices with ($b,~t$) exchange, that depends on the
specific combination $[m^2_t\cot^2\beta+m^2_b\tan^2\beta]$ (SUSY mass
parameters $M_i$, that could enter other vertex diagrams, would appear
in the form $\log(q^2/M^2_i)$, $\sqrt{q^2}$ is the c.m. energy, but with
a suitable change of scale they can always be shifted into a constant
term, as we shall show and discuss).\par
In Appendix we list the various logarithmic contributions coming from
different diagrams. The convention that we have followed is that of
keeping $M_W$ as the scale of the double $\log^2~q^2$ coming from two $W$
boxes and single $W$ vertices. For all the remaining logarithms we
have chosen a common scale $M_Z$, with the exception of the Yukawa
vertex where the most natural scale appears to us to be that of the
top mass $m_t$. This choice is arbitrary, it is only dictated by our
personal taste and feelings. The consequence of this choice will be
that of fixing the numerical value of the "next-to subleading" term. A
pragmatic attitude would be that of verifying that, with this choice,
\underline{both} the logarithmic and the next-to subleading terms 
remain acceptably "small" at the one loop level. 
This, as we shall show, will
happen in fact and with this "a posteriori" justification we shall
retain our choice of scales.\par
From a glance to the Appendix one also sees that the linear $\log~q^2$
belong to two separate classes: the so-called "universal" and the "not
universal" ones. The latter are produced by boxes and depend on the
c.m. scattering angle $\vartheta$. We have already discussed in a
previous paper \cite{scalar} their different relevance 
at the one loop level, and
we shall not insist here on these features. We can, though, recall the
observed fact \cite{scalar} that these $\vartheta$ dependent non universal terms are
the only ones that contribute at logarithmic level the forward-backward
asymmetries at one loop. Since they do not contain any SUSY parameter,
as we said, there will be no relevant logarithmic contribution
containing SUSY parameters to the forward-backward asymmetry of
$H^+H^-$ production. The same conclusion can be derived for the
longitudinal polarization asymmetry that is known to provide only
information on the initial state as shown in the previous reference
\cite{scalar}.  For this reason, we shall concentrate our
numerical analysis on the total cross section of the process.\par
After this rather technical presentation, we are now ready to present
the concrete numerical analysis. With this aim, we shall now summarize
our previous discussion giving the asymptotic relative effect on the
cross section of the process, by writing it in the form:

\bq
\Delta(q^2)={\sigma^{{\rm Born+1 loop}}-\sigma^{{\rm Born}}\over\sigma^{{\rm Born}}}
\eq\noindent
where in $\sigma^{{\rm Born+1 loop}}$ we are retaining only the genuine one
loop terms $\cal{O}(\alpha/\pi)$ and not the second order terms coming from
the square of the one loop contributions since these mix with the
genuine two loop contributions.\par
The logarithmic expansion of $\Delta$ has been derived analytically
and is given by the expression

\bqa
\Delta(q^2)&=& -~({\alpha\over2\pi s^2_W})
({1+2s^4_W\over1+4s^4_W})
\log^2{q^2\over M^2_W}\nonumber\\
&&-~({\alpha\over4\pi s^2_Wc^2_W})({1+2s^4_W+8s^6_W\over1+4s^4_W})
\log^2{q^2\over M^2_Z}\nonumber\\
&&-~({3\alpha\over4\pi s^2_WM^2_W})(m^2_t\cot^2\beta+m^2_b \tan^2\beta)
\log{q^2\over m^2_t}\nonumber\\
&&+~({\alpha\over3\pi s^2_Wc^2_W})
({11-16s^2_W+32s^4_W+72s^6_W\over1+4s^4_W})
\log{q^2\over M^2_Z}\nonumber\\
&&+\Delta_{\rm rem}(q^2)
\label{Delta}\eqa
\noindent
where the fourth line contains all single logarithms with the exception
of those of Yukawa origin (third line).
The last term $\Delta_{\rm rem}(q^2)$ is the difference
between the full one loop result and its asymptotic Sudakov expansion including all the double and single
logarithms, and will be called the next-to subleading term.

\section{Validity of the Sudakov expansion}

The aim of this Section is that of investigating whether there exists
a region of energy and of parameters where the rigorous calculation at
one loop can be reproduced by the effective Sudakov expansion
Eq.(\ref{Delta}), and to determine the relevant features of the
next-to subleading term $\Delta_{\rm rem}(q^2)$. In order to avoid
confusion, we anticipate that our analysis will be divided into two
different sectors. In the first one, summarized in this Section, we
shall study the dependence of $\Delta_{\rm rem}(q^2)$ on the c.m. energy
for given values of the parameters of the chosen MSSM model.
In the second one, we shall study the dependence of  
$\Delta_{\rm rem}(q^2)$ on the MSSM parameters for a fixed energy
chosen at the representative 1 TeV value.
To proceed with our analysis, we must now define the MSSM parameters
that we shall use as input of the calculations. 
We retained the following five free parameters:

\bq
\tan\beta,~~~\mu,~~~M_A,~~~M_2,~~~M_S
\eq
i.e. the ratio of the two vevs, the Higgs bilinear coupling, the CP
odd Higgs mass, the universal gaugino mass and the universal sfermion
mass. For this preliminary analysis, we allowed ourselves the
simplifications of using the GUT relation 
$M_1={5\over3}\tan^2\vartheta_W M_2$ and of setting the trilinear couplings
$A_u=A_d=0$. We computed the Higgs spectrum with the code FeynHiggsFast
\cite{FeynHiggsFast} and obtained the masses of charginos, neutralinos
and sfermions by numerical diagonalization of their mixing matrices.
We retained sfermion mixing only in the case of the third 
generation.\par
For the purposes of this paper, we have chosen to work in an energy
region between 800 GeV and 1 TeV, considering 1 TeV as an ambitious
conceivable final goal of the future LC. In our study, we shall assume
that a number of precise measurements can be performed in that energy
range, and we will show what could be the theoretical implications for
the model in a particular region of its couplings and of its masses.
Clearly, all the obtained informations and bounds on the various mass
parameters could be easily rescaled if the analysis were performed in
a lower (e.g. 600-800 GeV) energy interval.\par
The first problem that we address is that of determining the range of
massive MSSM parameters for which the Sudakov expansion  for
$\Delta(q^2)$ reproduces the rigorous calculation with a simple and
understandable expression for the next-to subleading term 
$\Delta_{\rm rem}(q^2)$. For the latter, the simplest possibility would be
provided by a constant, and from our previous SM experience
\cite{scalar}, we
would be prepared to the appearance in this case of a relatively
"large" (i.e. compared to the logarithmic component) quantity. We
cannot exclude, though, a priori the necessity of including an extra,
energy dependent term, that should vanish at very large $\sqrt{q^2}$,
 but could be numerically relevant and complicated in the
considered energy range. If this turned out to be the case, the
practical validity of the Sudakov expansion would be unavoidably
reduced, since every tentative effective fit to the data would be
complicated and affected by large errors, and we shall return on this
statement in the description of the derivation of $\tan\beta$, to be
shown in the forthcoming Section.\par
Given the fact that we have to deal with four massive parameters, we
have performed four different analyses, in each one of which three
parameters were fixed at values that we considered "light" with
respect to the chosen energy range, and one parameter was allowed to
vary. In all these analyses we fixed $\tan\beta$ at the value
$\tan\beta=20$, that can be considered as an average value in the range
that we have explored, roughly $2\lesssim \tan\beta\lesssim 40$ (we shall
discuss later on the reasons of this choice of the upper value).
Fixing different values of $\tan\beta$ does not change the results of
the four analyses, that we chose in the following way:

\begin{enumerate}
\item[a)] variable $\mu$, fixed $M_A = 200$ GeV, $M_2 = 100$ GeV, $M_S = 350$ GeV;
\item[b)] variable $M_A$, fixed $\mu = 300$ GeV, $M_2 = 100$ GeV, $M_S = 350$ GeV;
\item[c)] variable $M_2$, fixed $\mu = 300$ GeV, $M_A = 200$ GeV, $M_S = 350$ GeV;
\item[d)] variable $M_S$, fixed $\mu = 300$ GeV, $M_A = 200$ GeV, $M_2 = 100$ GeV;
\end{enumerate}

\begin{figure}
\centering
\epsfig{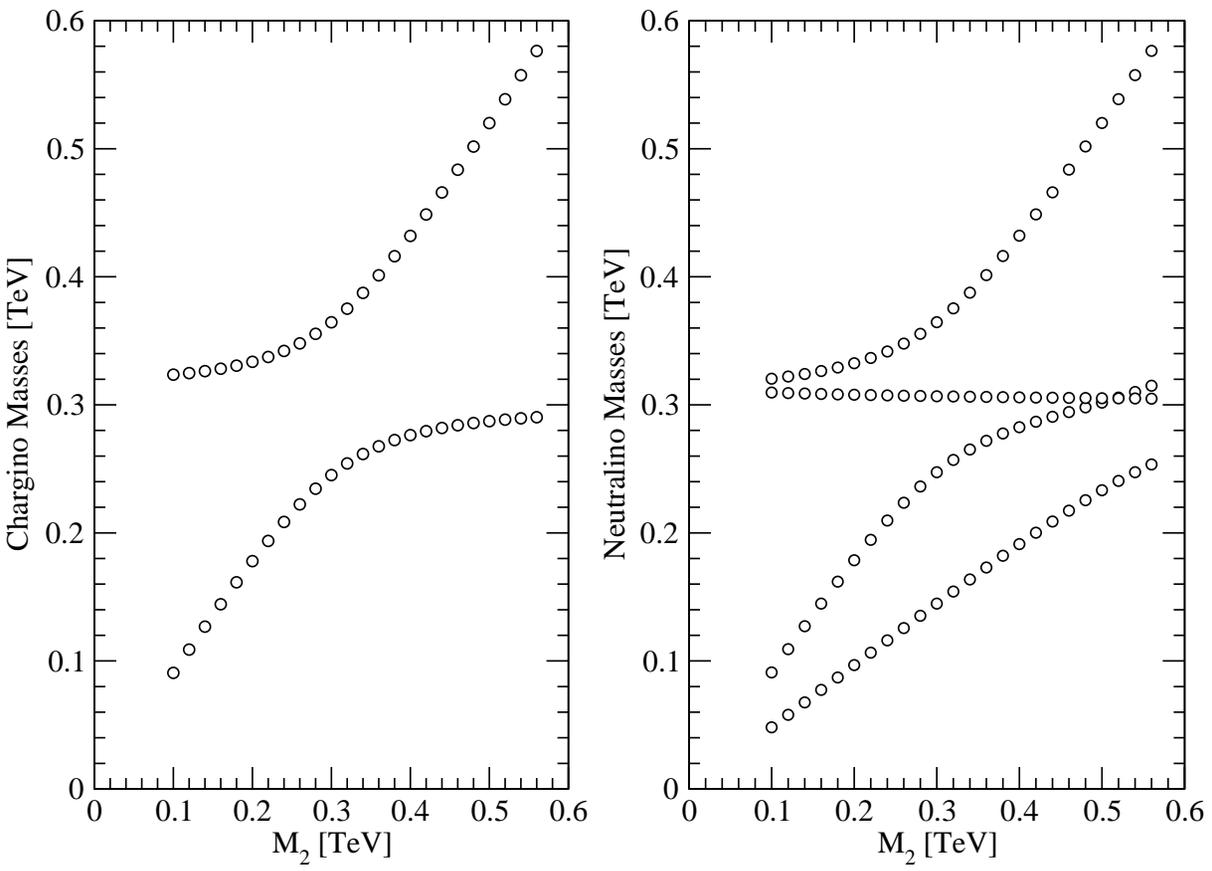}
\caption{Variable $M_2$: chargino and neutralino masses. Here $\tan\beta=20$.}
\label{varm2:gauginos}
\end{figure}

The choices of the fixed values of the parameters in the four analyses
are dictated by practical reasons and could be reasonably varied. The
relatively small value of $M_2$ corresponds to the request of having
relatively light charginos and neutralinos, although higher values
$(\lesssim 300~\mbox{GeV})$ for $M_2$ would still be acceptable, 
as shown  by Fig.~(\ref{varm2:gauginos}). For $M_A$, 
the same considerations apply in order to avoid
resonance formation. $M_S$ can vary in a larger range
without apparent
problems, and therefore it was fixed at a relatively higher value.\par
We are now ready to discuss the results of our analyses. This was
performed by computing numerically the quantity $\Delta(q^2)$ with the
specific code (SESAMO) that we have built for our purposes and
which is, we repeat, available for use. From the computed quantity we
then subtracted all the logarithms of Eq.(\ref{Delta}) and obtained
the remaining, next-to subleading term $\Delta_{\rm rem}(q^2)$.\par
An important preliminary remark must be made at this point. The final
goal of our analysis is that of showing that a complete one-loop
expression can be reproduced by a (relatively simple) Sudakov
expansion. But a necessary condition to make this search useful is also
that the one-loop expansion is reliable. To guarantee this condition,
we shall have to verify that the complete expression of $\Delta(q^2)$
(i.e. the full one-loop correction) remains acceptably small in the
investigated energy region, with e.g. an upper bound that we could fix
quantitatively at the relative ten percent value. Only after this
check, our next analysis could be considered as meaningful.\par
We are now ready to show the results of the four considered cases,
which are the following ones:\\

1) Case (a): variable $\mu$\\

\begin{figure}
\centering
\epsfig{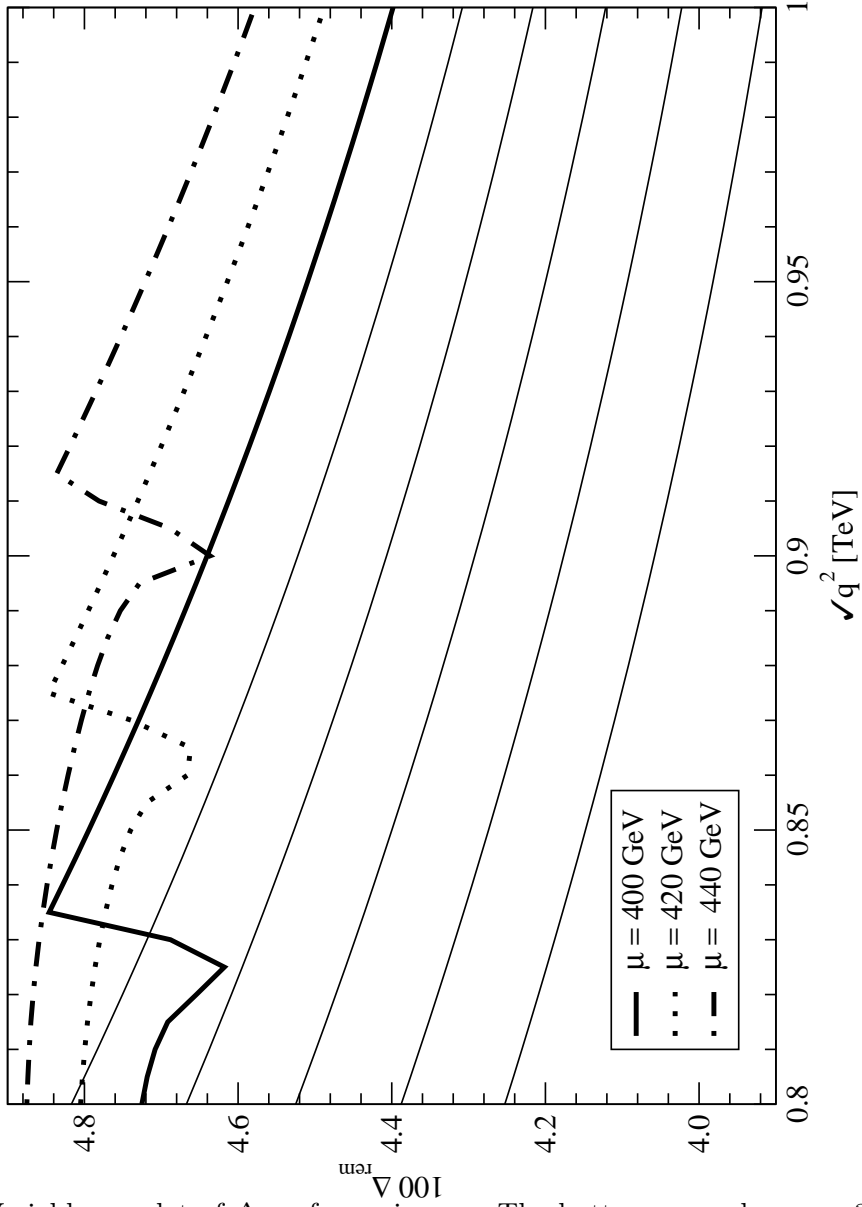}
\caption{Variable $\mu$: plot of $\Delta_{\rm rem}$ for various $\mu$. The bottom curve has
$\mu = 300$ GeV. The other curves have increasing $\mu$ by steps of 20 GeV.}
\label{varmu:delta}
\end{figure}

We allowed $\mu$ to vary from an initial value of 300 GeV
to final values of approximately 400 GeV (below 300 GeV, we encountered
problems in the determination of the sfermion mass eigenstates whose
discussion seems to us beyond the purposes of this preliminary analysis).
The full one-loop effect,
computed at the representative value $\sqrt{q^2}=1~\mbox{TeV}$, remains
always very small (below two percent). The values of  
$\Delta_{\rm rem}(q^2)$ in the interval $800~\mbox{GeV}\leq \sqrt{q^2}\leq 1~\mbox{TeV}$
are shown in Fig.~(\ref{varmu:delta}). One sees that the term is changing by an amount
roughly equal to five percent of its central value when moving from
the beginning to the end of the interval. Numerically, this corresponds
to a two permille effect that can be considered as "fairly small"
under our expected experimental conditions of a one percent accuracy.
When $\mu$ becomes larger than roughly $380~\mbox{GeV}$, the simplicity of 
$\Delta_{\rm rem}(q^2)$ is lost and a complicated energy dependence
appears which is due to a resonance effect:
when we increase
$\mu$, one of the two charginos and two neutralinos 
become progressively heavier with masses $\sim \mu$.
When $\mu \sim 400$ GeV, we begin to see a kinematical 
threshold at $\sim 800$ GeV that produces a bump 
in $\Delta_{\rm rem}$
well visible in the figure. Of course, the bump shifts 
to the right as $\mu$ is further increases.
One observes also from Fig.~(\ref{varmu:delta}) that $\Delta_{\rm rem}(q^2)$ is scarcely
affected by variations in $\mu$ (about half percent when $\mu$ varies
from $300$ to $400$ GeV) and, as a consequence, it does not appear to
be a promising candidate for testing virtual MSSM effects generated by
this specific parameter.\par
We repeated our analysis changing only the sign of $\mu$. The obtained
curves are practically (in agreement with our previous statement)
identical with those corresponding to positive $\mu$ values. For this
reason, we shall restrict ourselves from now on to considering
conventionally the $\mu>0$ scenario.\\

2) Case (b): variable $M_A$\\

\begin{figure}
\centering
\epsfig{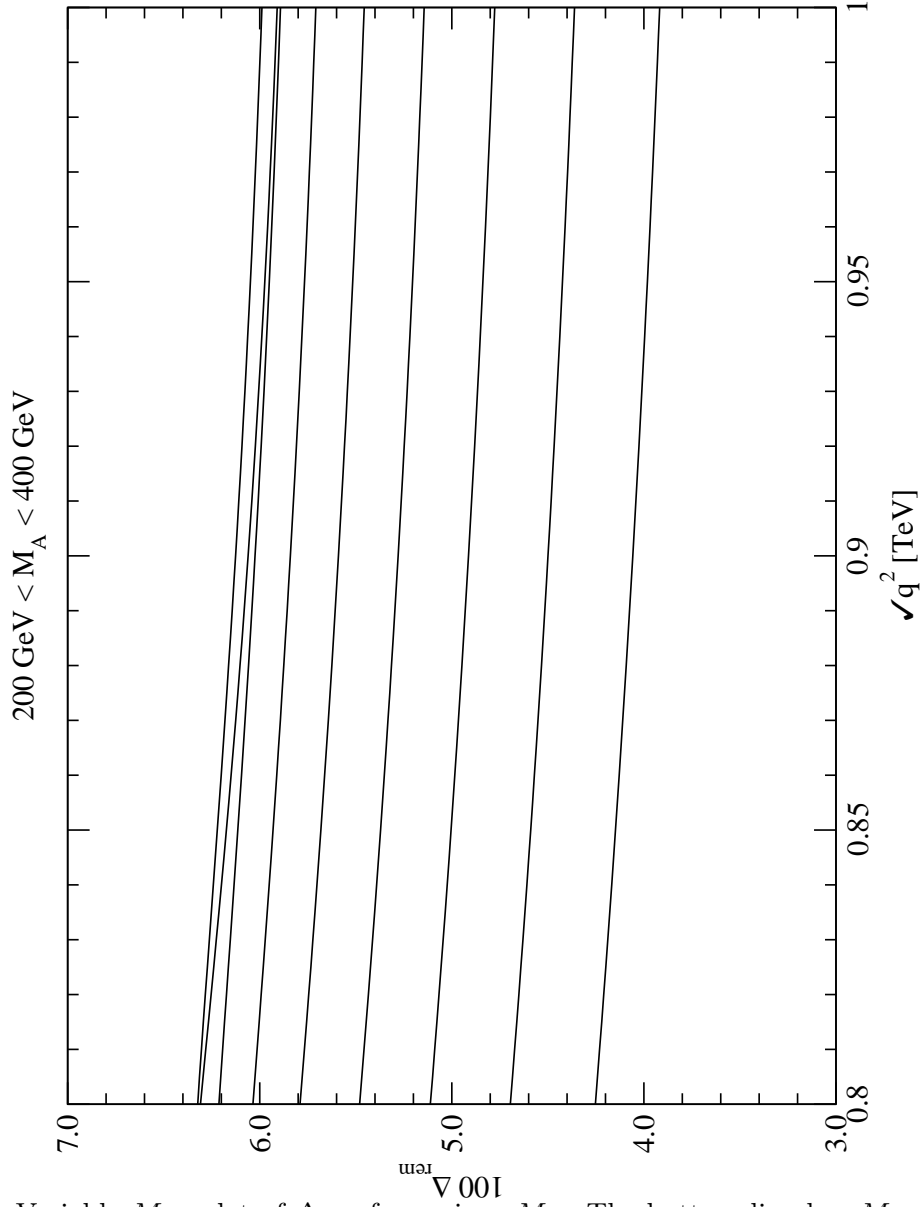}
\caption{Variable $M_A$: plot of $\Delta_{\rm rem}$ for various $M_A$. The bottom line has
$M_A = 200$ GeV. The other curves have increasing $M_A$ by steps of 20 GeV.}
\label{varma:delta}
\end{figure}

We have varied $M_A$ starting from an initial value of $200~\mbox{GeV}$ up to
approximately $400~\mbox{GeV}$. The full (negative)
one-loop effect at $1~\mbox{TeV}$ remains
systematically below a relative two percent.
As $M_A$ increases, the basic change in the spectrum of SUSY 
particles is that both $H^+$ and $H^0$
become heavier with approximately $M_{A^0}\sim M_{H^0}\sim M_{H^+}$. 
This means that in the plots 
we have to take into account the kinematical constraint 
$\sqrt{q^2}\ge 2M_{H^+}$. In  
Fig.~(\ref{varma:delta}) we show the behaviour of 
$\Delta_{\rm rem}$ for $200~ \mbox{GeV}< M_A < 400~\mbox{GeV}$.
Once again, we notice that in the considered energy interval
$\Delta_{\rm rem}(q^2)$ remains "essentially" constant, with relative
extreme variations of five percent (or less) from its central value.
This simple pattern would be lost for larger $M_A$ values $>400~\mbox{GeV}$,
due to the aforementioned Higgs production threshold. One notices in
this case that the dependence of $\Delta_{\rm rem}(q^2)$ on $M_A$ is
sizable: to a variation of $M_A$ from $200$ to $300~ \mbox{GeV}$ there
corresponds a variation in $\Delta_{\rm rem}(q^2)$ of almost two percent,
that would be visible at the proposed LC.\\

3) Case (c): variable $M_2$\\

\begin{figure}
\centering
\epsfig{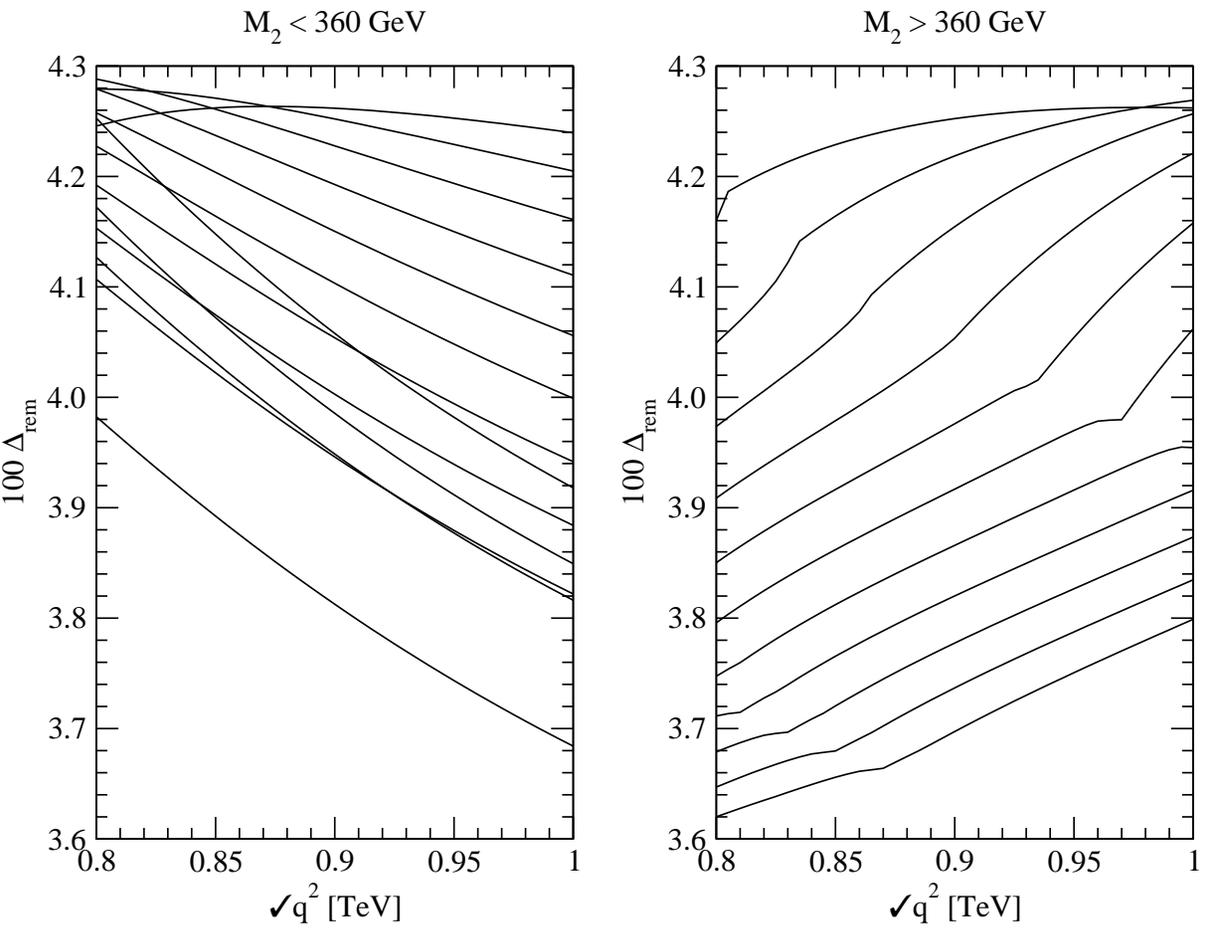}
\caption{Variable $M_2$: plot of $\Delta_{\rm rem}$ for various 
$M_2$ between 100 and 560 GeV. The various 
curves have values of $M_2$ spaced by 20 GeV.}
\label{varm2:delta}
\end{figure}

Here, we vary $M_2$ from $100~\mbox{GeV}$ to $400~\mbox{GeV}$. The full one-loop
effect at $1~\mbox{TeV}$ is again very small (below two percent) and negative.
In the considered range of variation of $M_2$, there are one 
chargino and one neutralino with masses
increasing approximately as $M_2$ and reaching the $400$ GeV 
value at which some resonance structure
can be observed in the plots of $\Delta_{\rm rem}$ shown in
Fig.~(\ref{varm2:delta}). It can 
be noticed that, 
for $M_2$ larger than about $360$ GeV, the shape of 
this function is definitely not constant with energy and clear 
bumps can be seen, 
due to resonance effects associated to the heavy gauginos.
As in the case of $\mu$, we observe that $\Delta_{\rm rem}(q^2)$
has very small sensitivity to the variations of $M_2$ (about two
permille variation for a $100~\mbox{GeV}$ shift in $M_2$).\\

4)  Case (d):  variable $M_S$\\

\begin{figure}
\centering
\epsfig{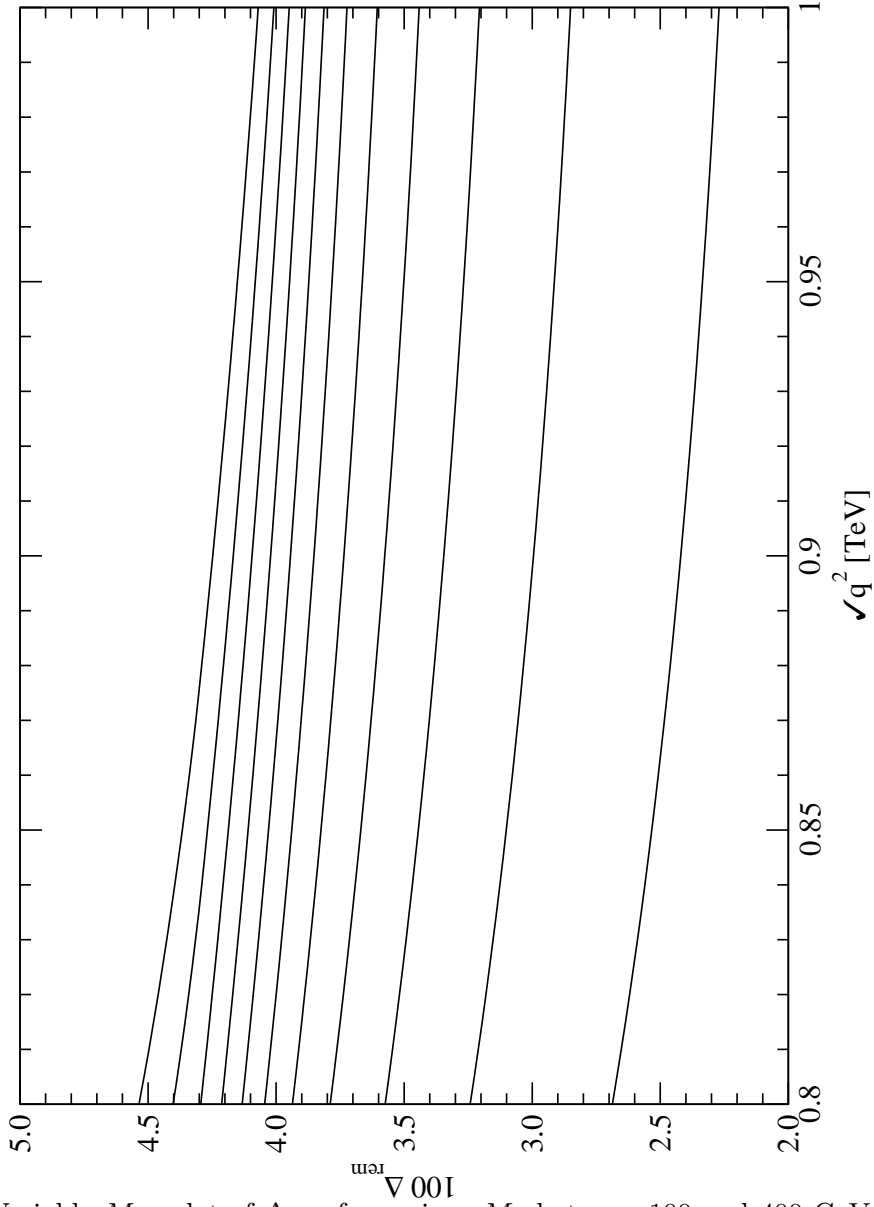}
\caption{Variable $M_S$: plot of $\Delta_{\rm rem}$ for 
various $M_S$ between 100 and 400 GeV. The 
bottom line has $M_S = 100$ GeV. The other curves have increasing $M_S$ by
steps of 20 GeV.}
\label{varms:delta}
\end{figure}

We have varied $M_S$ from $100~\mbox{GeV}$ to about $400~\mbox{GeV}$. The full
one-loop effect is again very small and negative
(below $\simeq3-4$ percent). The
values of $\Delta_{\rm rem}(q^2)$ shown in Fig.~(\ref{varms:delta}) are again essentially (up
to a relative $5$ percent) constant in the considered energy interval.
One sees that $\Delta_{\rm rem}(q^2)$ is sensitive to $M_S$: $100~\mbox{GeV}$ of
variation in this parameter corresponds to $\simeq1.5$ percent
variation in $\Delta_{\rm rem}(q^2)$.\\

To summarize, we have considered the range of validity of a
logarithmic Sudakov expansion with an extra next-to subleading term
which is "essentially" constant in a certain domain of the massive
MSSM parameters, for c.m. energies in the range $800~\mbox{GeV}-1~\mbox{TeV}$. In
our approach, to be "essentially" constant means to be well
approximated by the central value, with a few permille error. We have
verified that this requirement is met for values of all the parameters
below, approximately, $350~\mbox{GeV}$. This gives a quantitative
illustration of how "light" the MSSM parameters should be in order
that a typically asymptotic expansion like the Sudakov one that we
investigate might hold at energies in the $1~\mbox{TeV}$ range. We can remark
that, as matter of fact, the validity of the expansion corresponds to
values of the masses meeting the naive request $M^2_i/q^2\lesssim
10^{-1}$. This remains true even if several resonances appear close to
$\sqrt{q^2}/2$.
In other words the energy dependence of this process appears to be
reasonably flat in the considered domain. We also conclude that the
values of the "essentially" constant components are relatively "large"
and of opposite sign (positive) with respect to the (negative)
logarithmic contribution (roughly, they are of comparable size,
although always smaller than the logarithmic ones). As a result of
this cancellation the overall one-loop effect is sensibly reduced,
becoming systematically of few percent size. 
This situation reproduces exactly the one
that we met in the analogous study of the SM case, even if the
number of free parameters in the MSSM is so much larger.\par
A final question that we asked ourselves was that of whether the
approximate constancy of $\Delta_{\rm rem}(q^2)$ could be due to the
relative smallness of the investigated energy domain. To answer this
question, we have extended our analysis to very large $\sqrt{q^2}$
values up to $10~\mbox{TeV}$, for an illustrative fixed set of MSSM
parameters

\bq
\tan\beta = 30,\ \mu = 300,\ M_A = 200,\ M_2 = 100,\ M_S = 350,
\eq

\begin{figure}
\centering
\epsfig{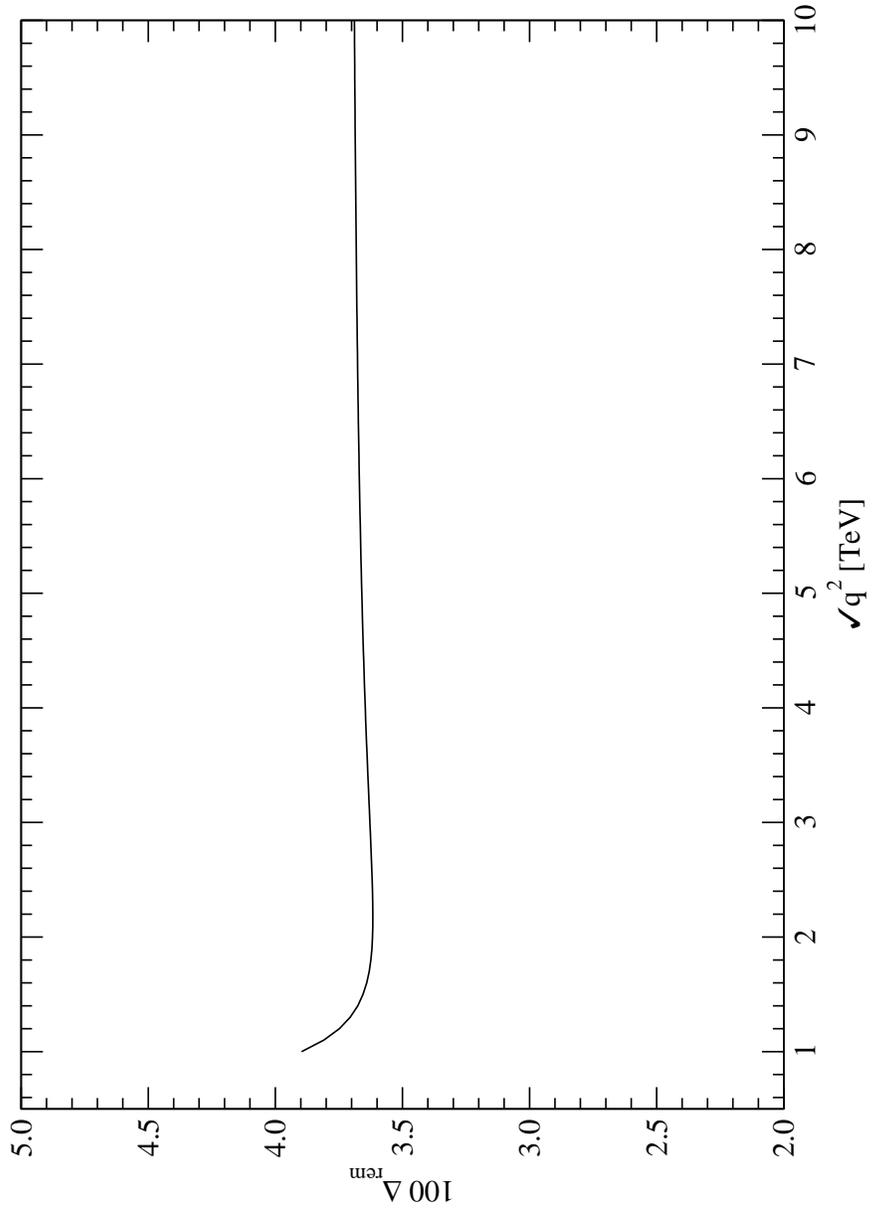}
\caption{Behaviour of $\Delta_{\rm rem}$ in the ultra high energy regime.}
\label{UHE}
\end{figure}

The results are shown in Fig.~(\ref{UHE}). One sees that for energies beyond
$\simeq2~\mbox{TeV}$ the values of $\Delta_{\rm rem}(q^2)$ remain practically
(i.e. to less than one permille) constant. Although we already know
that at such energies the one-loop approximation is probably not
valid, we consider this result as a check of the asymptotic validity
of the expansion (and also of the numerical code that 
we have used).\par
In conclusion, we have seen that in the $1~\mbox{TeV}$ range a 
1-loop Sudakov
expansion with an "essentially" constant term is valid for the
defined "light" SUSY scenario. The fact that $\Delta_{\rm rem}(q^2)$
is not rigorously constant is expected to produce therefore
reasonably "small"
effects, that we shall try to evidentiate in full detail in the
forthcoming Section 4.

\section{Study of the MSSM parameters}

\subsection{Determination of $\tan\beta$}

The first question that we address is the relevance of 
the Sudakov expansion for the problem
of determining $\tan\beta$ under the assumption that the 
mass scales are in the ``safe'' range 
that we have just discussed. In other words, 
$\Delta_{\rm rem}$ is only approximately constant and we 
want to understand quantitatively how much this 
can affect an attempt to determine $\tan\beta$ from the 
logarithmic slope of the cross-section, in which it appears in the
combination shown in Eq.(\ref{Delta}).\par 
With this aim, we begin by subtracting explicitly from 
$\Delta$ all the ``known'' logarithms, 
i.e. all the terms in the Sudakov expansion with the exception 
of the Yukawa contribution. 
Therefore, we define the quantity
\bqa
\widetilde\Delta(q^2) &=& F(\tan\beta) \log\ q^2 + 
\Delta_{\rm rem}(q^2) ,\\
F(\tan\beta) &\equiv& -\frac{3\alpha}{4\pi s_{\rm w}^2 M_W^2}
(m_t^2\cot^2\beta+m_b^2\tan^2\beta)\nonumber .
\eqa
If, in a definite scenario, the shape of 
$\Delta_{\rm rem}(q^2)$ turned out to be flat, 
then it would be conceivable to approximate it by a 
term that is constant with respect to $q^2$. 
In such a simple case, we could try to fit the measured values 
of the residual effect $\widetilde\Delta(q^2)$ with 
a logarithmic expansion in $q^2$ of the form 
\bq
A_{\rm fit}\log q^2 + B_{\rm fit} .
\eq 
The result of the fit, $A_{\rm fit}$, can be compared with $F$ at the value of 
$\tan\beta$ we are working. The difference $\delta F = A_{\rm fit}-F$ is an error in the estimate of 
$F$ that has two components: $\delta F = \delta_{\rm stat}F + \delta_{\rm sys} F$.
The first term, $\delta_{\rm stat}F$ is simply due to the fact that we assume a 
certain finite experimental precision on each measurement. The second term, 
$\delta_{\rm sys}$ is the most important and is a systematic error due to the 
fact that $\Delta_{\rm rem}$ is not constant with respect to the energy. 
For instance, if $\Delta_{\rm rem}$ were exactly energy independent, we would find 
$\delta_{\rm sys} F = 0$.
The error $\delta F$ can be converted into an error on the  
estimate of the interesting parameter $\tan\beta$. 
If $\delta F/F$ is enough narrow to allow a linearized analysis, then we have simply
\bq
\frac{\delta F}{F} = \frac{\tan\beta\ F'}{F}
\ \frac{\delta\tan\beta}{\tan\beta}
\eq
or
\bq
\frac{\delta\tan\beta}{\tan\beta} = \frac{1}{2}
\ \frac{\tan^4\beta + (m_t/m_b)^2}{\tan^4\beta-(m_t/m_b)^2}
\ \frac{\delta F}{F}
\eq
The zero in the denominator corresponds to the value 
$\tan\beta = \sqrt{m_t/m_b}\simeq 6.2$ at which 
the function $F$ attains its maximum and the sensitivity 
to $\tan\beta$ is the smallest due to the 
flatness of $F$. Notice also that for $\tan\beta$ beyond 
$15-20$, we have 
\bq
\frac{\delta F}{F} \simeq \frac{1}{2}
\ \frac{\delta\tan\beta}{\tan\beta}
\eq
In the following discussion we shall analyze in a quantitative way the feasibility of such a 
procedure in the framework of specific scenarios. With this aim, we have assumed the existence of 
10 equally spaced experimental measurements in the range $800~\mbox{GeV}-1~\mbox{TeV}$ with 
a relative 1 percent precision and have generated them by means of our numerical code
(if only $N$ points are available, all the numerical results concerning the statistical component 
of the error on $\tan\beta$ must be increased by a factor $\sqrt{10/N}$).
\\

\underline{L: Very Light SUSY}

\begin{figure}
\centering
\epsfig{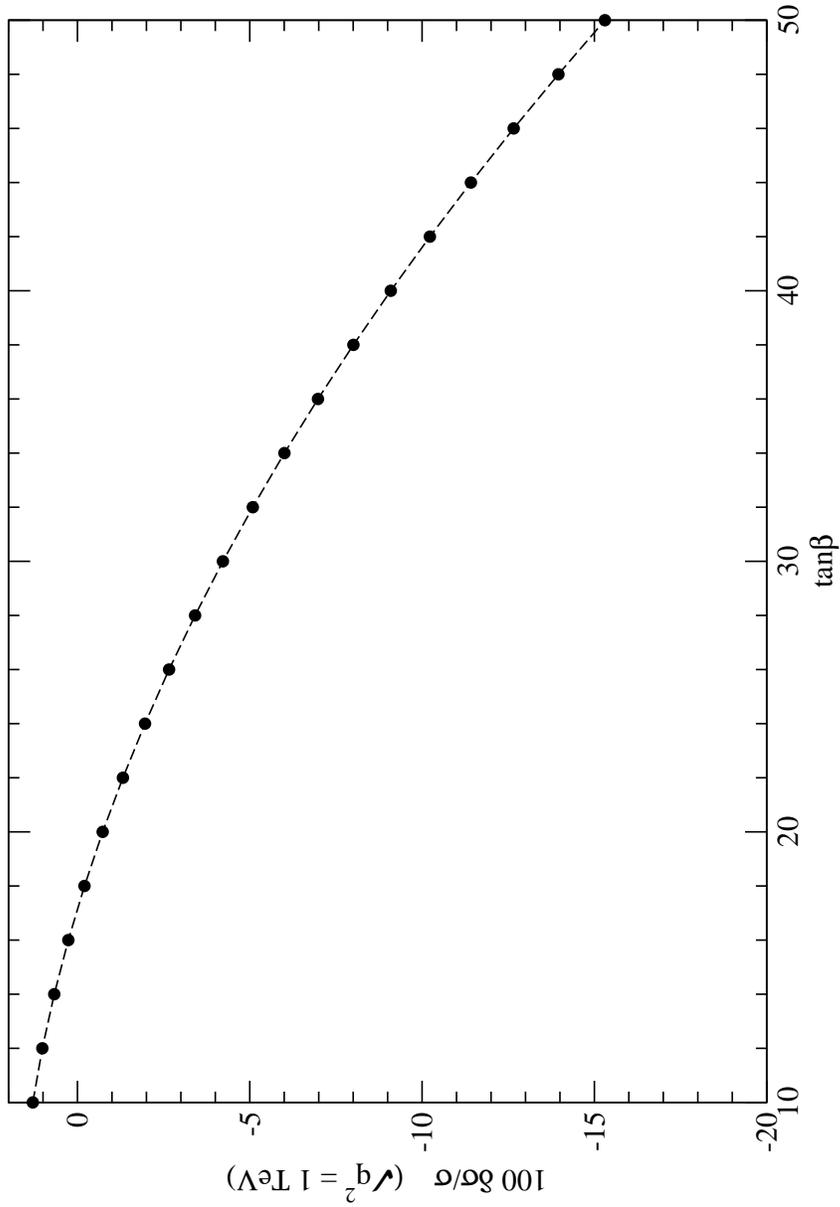}
\caption{Variable $\tan\beta$ in the (L) scenario: full effect at 1 TeV}
\label{varbeta:L:effect}
\end{figure}

\begin{figure}
\centering
\epsfig{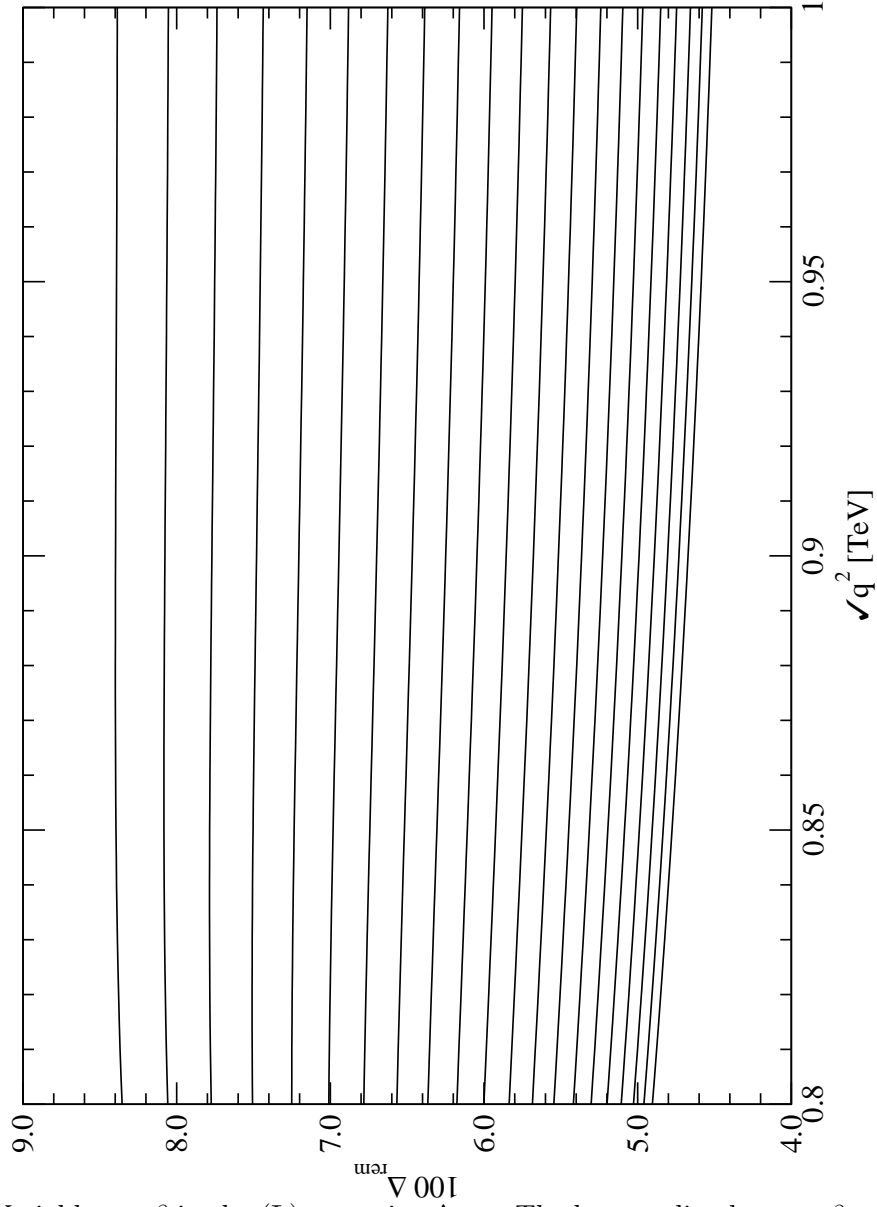}
\caption{Variable $\tan\beta$ in the (L) scenario:  $\Delta_{\rm rem}$. The bottom line
has $\tan\beta=10$, the other curves have increasing $\tan\beta$ by steps $\Delta\tan\beta=2$.}
\label{varbeta:L:delta}
\end{figure}

\begin{figure}
\centering
\epsfig{file=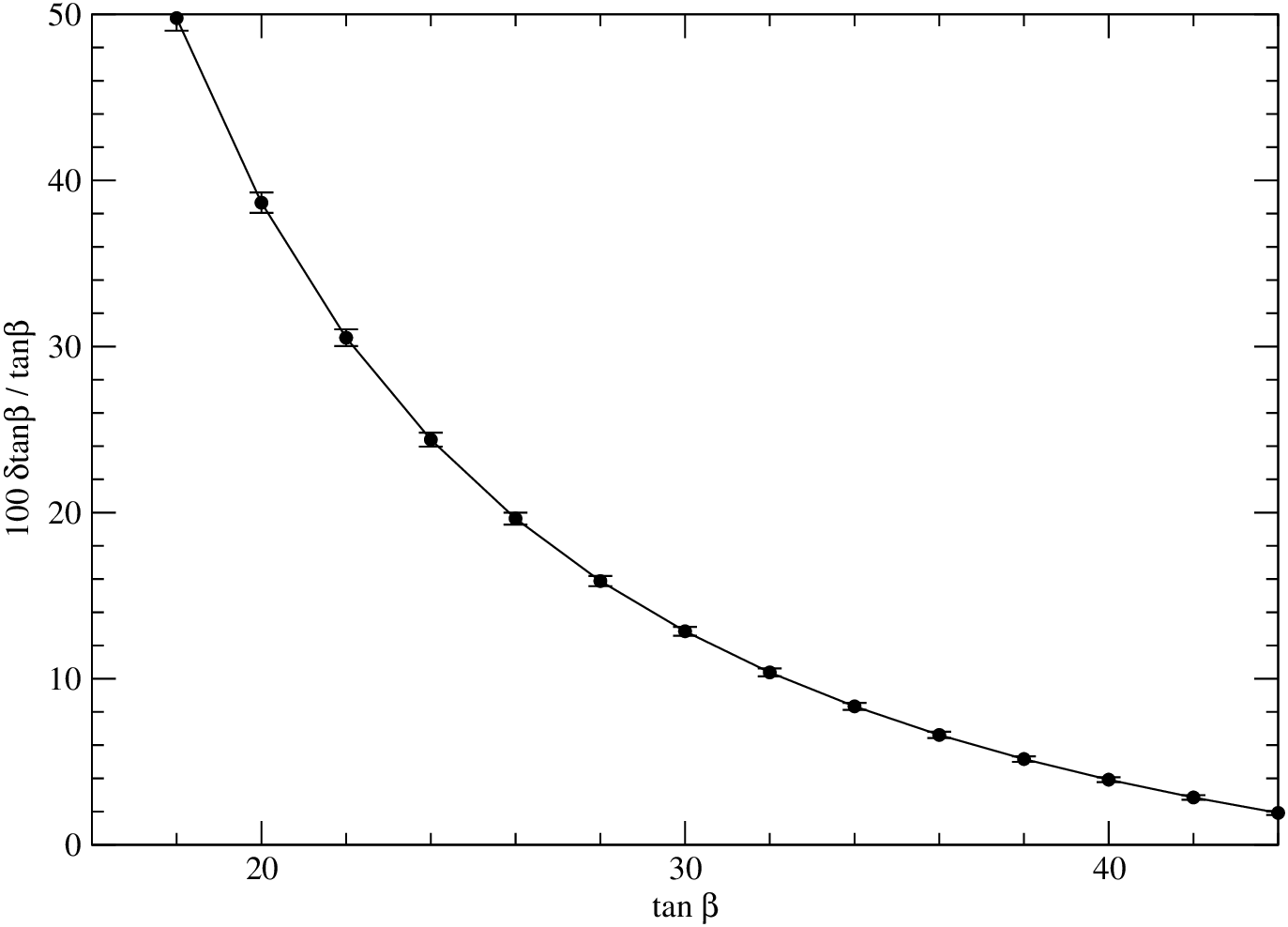,width=16cm,angle=90}
\caption{Variable $\tan\beta$ in the (L) scenario: 
percentual relative error in the determination of $\tan\beta$ at various 
$\tan\beta$.}
\label{varbeta:L:error}
\end{figure}

\bq
(L)\ \mbox{variable}\ \tan\beta\qquad
\mu = 300,\ M_A = 250,\ M_2 = 100,\ M_S = 350 .
\eq
The full effect at 1 TeV is given in  Fig.~(\ref{varbeta:L:effect}).
showing that it remains below a 10 percent for $\tan\beta\lesssim40$.
The curves for $\Delta_{\rm rem}$ are given in 
Fig.~(\ref{varbeta:L:delta}) showing that $\Delta_{\rm rem}(q^2)$
depends effectively on $\tan\beta$, remaining "essentially" constant in
the considered energy range.
The plot of the relative error in the determination of 
$\tan\beta$ is shown in 
Fig.~(\ref{varbeta:L:error}). As we said, the extra error bars are 
due to the fact that $A$ is determined with 
a statistical error due to the assumed 1\% accuracy in 
the cross section measurements. One sees that the main source of error
is actually due to the departure of $\Delta_{\rm rem}(q^2)$ from its
constant value.\par
We have subsequently considered two more scenarios, defined as:\\

\underline{ A: Light SUSY}

Here, we increase the masses in the gaugino sector.
\bq
(A)\ \mbox{variable}\ \tan\beta\qquad
\mu = 300,\ M_A = 250,\ M_2 = 200,\ M_S = 350 .
\eq

\underline{ B: Light SUSY with larger $\mu$}

\bq
(B)\ \mbox{variable}\ \tan\beta\qquad
\mu = 400,\ M_A = 250,\ M_2 = 200,\ M_S = 350 .
\eq

\begin{figure}
\centering
\epsfig{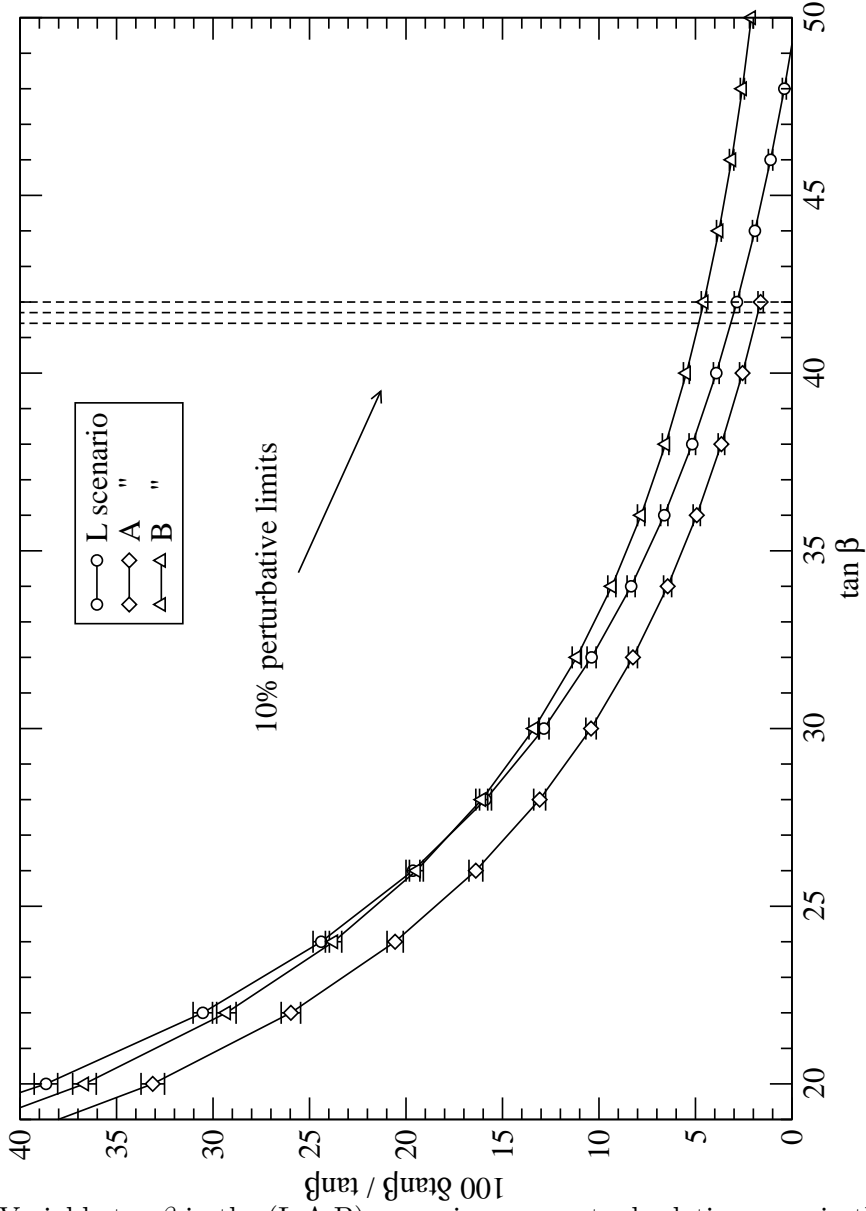}
\caption{Variable $\tan\beta$ in the (L,A,B) scenarios: 
percentual relative error in the determination of $\tan\beta$ at various 
$\tan\beta$.}
\label{varbeta:combined:error}
\end{figure}

In Fig.~(\ref{varbeta:combined:error}) we combine the 
results for the various
scenarios. In the figure, we have also shown the 
vertical lines corresponding to the 
{\em safe} perturbative bound corresponding to a 
10\% full one loop effect. One can see that, for $\tan\beta$ larger than
$20$, a determination of this parameter to better than a relative $40$
percent would be possible. For values larger than $\simeq30$, the
error would be reduced below a remarkable $\simeq10$ percent limit. 
This would remain true even in the worst case of relatively heavy SUSY
scenario (3) with e.g. $M_A=250~\mbox{GeV}$, and would represent to
our knowledge a valuable possibility of determining this fundamental
MSSM parameter in the region of high values where it is known
\cite{Datta} that accurate measurements are rather difficult.

\subsection{Visible effect of the remaining parameters}

Our logical scheme for extracting information from charged Higgs
production would now proceed in the following way. Once the proposed
determination of $\tan\beta$ from measurements of the slope of the cross
section were completed, we would return to the remaining term
$\Delta_{\rm rem}(q^2)$ and estimate the effects on it of the remaining
parameters assuming a precise measurement at a fixed energy, typically
$1~\mbox{TeV}$. A preliminary request will be that of taking into account the
error on the previous determination of $\tan\beta$. Following the
illustrations of Section 3, we shall optimistically assume that
$\tan\beta$ has been determined at a "convenient" value, i.e. one where
the relative error is of the ten percent size. For purposes of
illustration, we shall chose $\tan\beta=30\pm3$ from now on. Using this
value as a given input, we can examine which information on the
remaining parameters can be obtained from the determination of
$\Delta_{\rm rem}$. This determination will be affected by two sources: a
purely experimental one from the measurement at $1~\mbox{TeV}$, treated under
the usual assumptions, and the input error on $\tan\beta$ measured from
the slope. It is not difficult to see (e.g. looking at Fig.~(\ref{varbeta:L:delta}) and
considering Eq.(\ref{Delta})) that the latter will affect the
determination of $\Delta_{\rm rem}$ by a tolerably small (few permille)
error. We shall take it into account in what follows within
qualitative limits, not to make this indicative treatment too
involved.\par
\begin{figure}
\centering
\epsfig{file=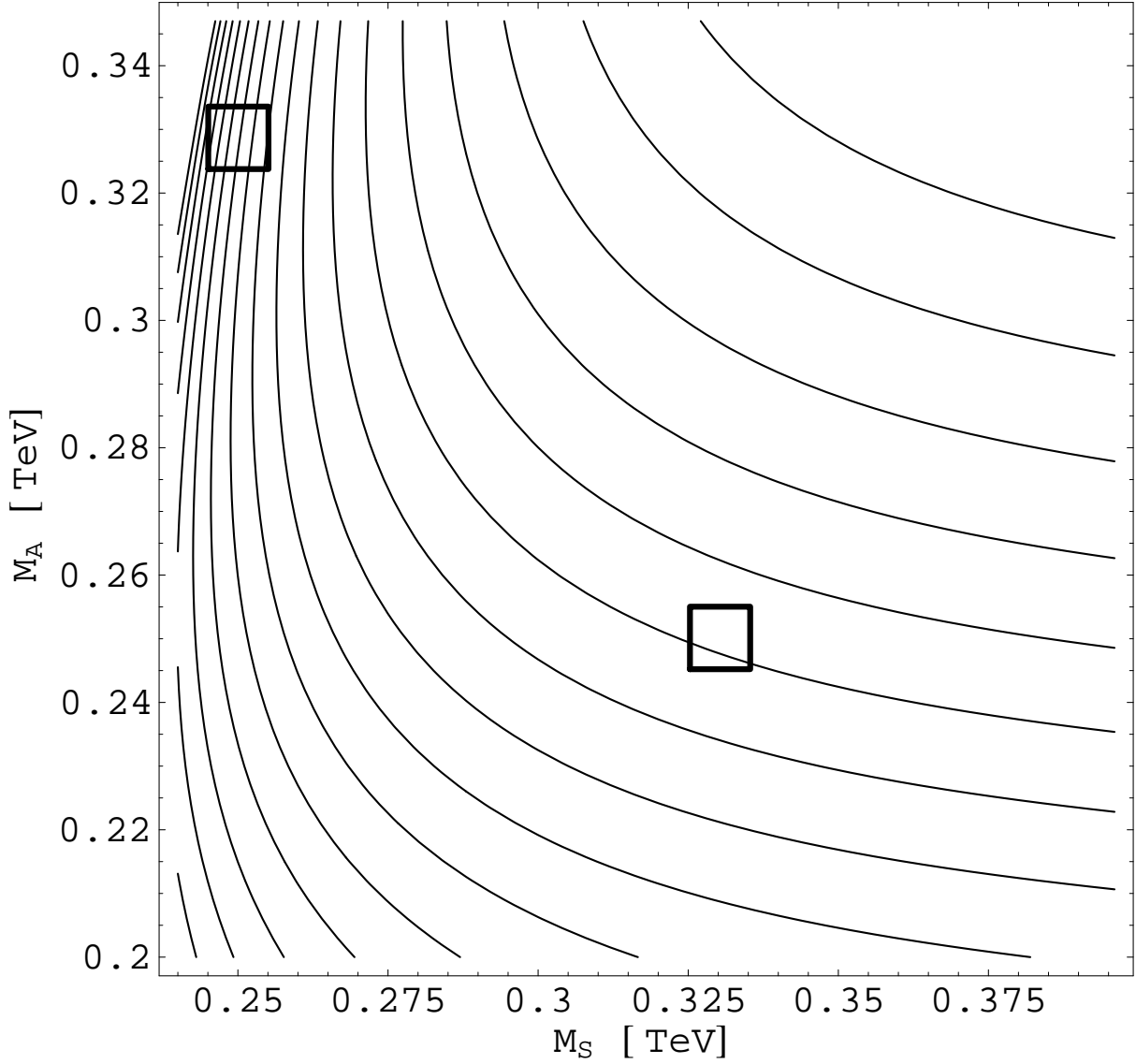 ,width=16cm}
\vskip 1cm\caption{Contour plot of $100\Delta_{\rm rem}(1 TeV)$ in the plane $(M_A, M_S)$.
The contour lines correspond to values between 0 and 8.5 increasing 
from left to right by steps of 0.5. Also shown are the boxes corresponding
to the points $(M_A, M_S) = (250\pm 5\ \mbox{GeV}, 330\pm 5\ \mbox{GeV})$ and 
$(M_A, M_S) = (330\pm 5\ \mbox{GeV}, 250\pm 5\ \mbox{GeV})$.}
\label{mamscontour}
\end{figure}
\begin{figure}
\centering
\epsfig{file=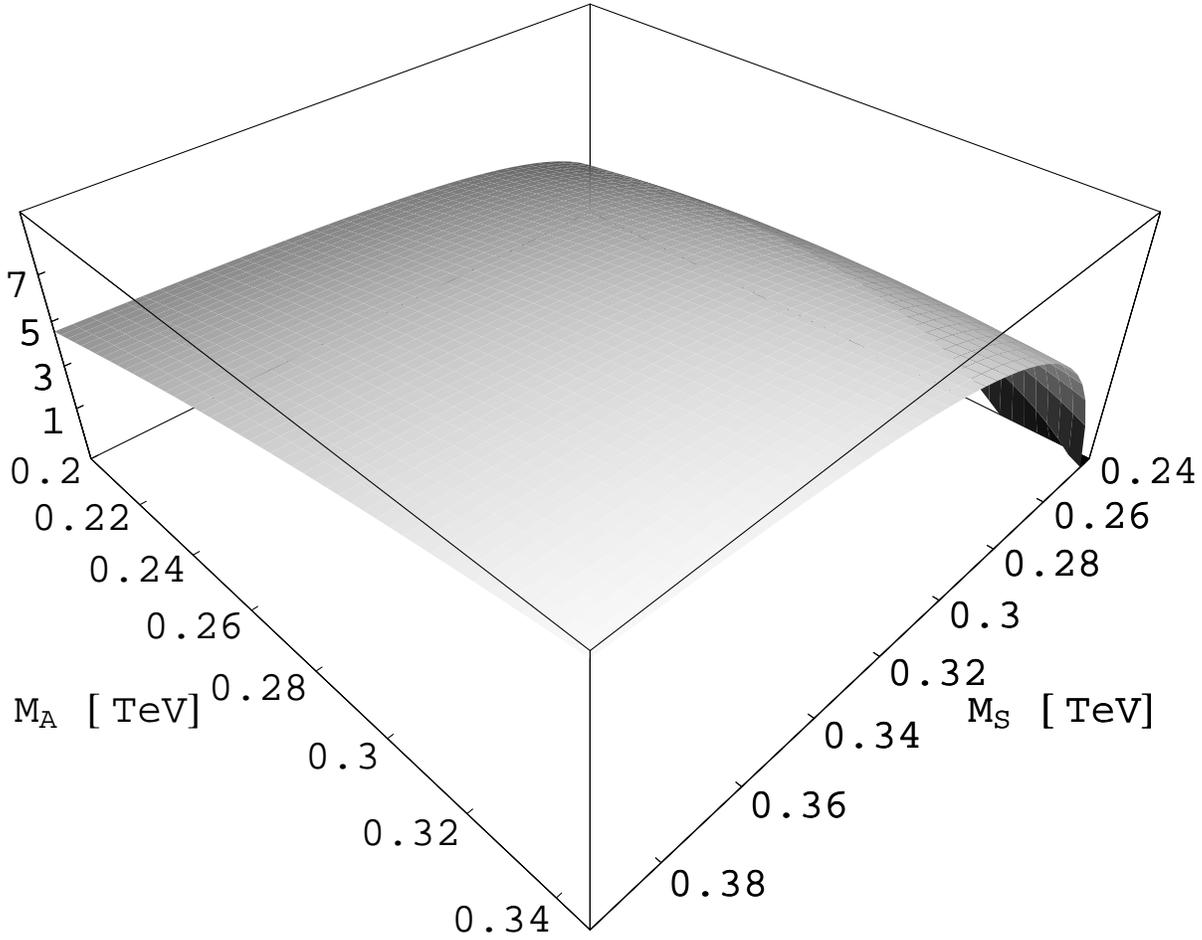,width=16cm}
\vskip 1cm\caption{Surface plot of $100\Delta_{\rm rem}$ at 1 TeV in the plane $(M_A, M_S)$.}
\label{mamssurface}
\end{figure}

The plan of our forthcoming study has been remarkably helped by the
observation that we already made, which shows that, in practice, 
$\Delta_{\rm rem}$ remains "essentially" unaffected by variations of $\mu$
and of $M_2$ in the considered "light" scenario. This simplifies
our approach, reducing it to the "essential" parameters, that are
$M_A$ and $M_S$. We have thus fixed $\mu,~M_2$ at conventional values
($\mu=400~\mbox{GeV}$, $M_2=100~\mbox{GeV}$) and drawn the contour and the surface
plots in the ($M_S, ~M_A$) variables shown by
Figs.\ref{mamscontour},\ref{mamssurface}.\par
A few, necessarily qualitative, comments are now appropriate e.g. from
a glance to Fig.\ref{mamscontour}. The various curves 
correspond to variations of 
$\Delta_{\rm rem}$ at 1 TeV. The spacing between two curves is a shift in 
$\Delta_{\rm rem}$ of 5 permille, which corresponds roughly to one half or
our expected error on this quantity and, in our Figure, defines a certain
bi-dimensional "tube" whose slope and width depend on the parameter
domain and fix the corresponding domain bounds on ($M_A,~M_S$). One
notices that, independently of the value of $\Delta_{\rm rem}$, there
would be a kind of orthogonal situation. For small ($\lesssim 4$
percent) $\Delta_{\rm rem}$ values, one would feel the effect of $M_S$
with a certain accuracy (of about $50~\mbox{GeV}$) without practical effect
from $M_A$. For larger $\Delta_{\rm rem}$ values, the opposite situation
would appear, and effects of $M_A$ could be felt to the previous
(about $50~\mbox{GeV}$) accuracy. These accuracies are certainly much worse
than the expected precisions on $M_A$, $M_S$ from direct production
(roughly, a relative 1-2 percent). However, in our opinion, these
curves could still be rather meaningful for a possible non trivial
consistency test of the model. Assuming in fact that both $M_A$ and
$M_S$ have been determined in a range between $200$ and $350~\mbox{GeV}$ with
a precision of, say, five GeV, the point in the ($M_A,~M_S$) plane
that corresponds to these values must lie on the "correct" curve that
corresponds to the measured value of $\Delta_{\rm rem}$. In
Fig.\ref{mamscontour} we have
drawn for illustration purposes two points that correspond to typical
couples of "light" values $M_A=250~\mbox{GeV},~M_S=330~\mbox{GeV}$ and
$M_A=330~\mbox{GeV},~M_S=250~\mbox{GeV}$ with the corresponding assumed experimental
error. One sees that a measurement of $\Delta$ to the relative
one percent accuracy would be of scarce use for 
$M_A=330~\mbox{GeV},~M_S=250~\mbox{GeV}$, but would provide a quite stringent test
of the model for the symmetrical couple of values. 
Thus, depending on the experimental results on
these masses, the relevance and the motivations of the previous
analysis at one TeV might become definitely enhanced.

\section{Conclusions}

The main conclusions that may be drawn from our analysis of the
charged Higgs production process are, in our opinion, the following\\

1) For this process, in the $\simeq$ 1 TeV energy region, an
effective one-loop description with a Sudakov expansion implemented by an
"essentially" constant next-to subleading term reproduces the rigorous
calculation in a "light" SUSY scenario where all the relevant mass
parameters of the process are roughly below the common $350~\mbox{GeV}$
value. The overall one-loop effect remains systematically under control
(below a safe few percent limit) in this region and seems to provide a 
reliable description of the process.
\\

2) A satisfactory determination of $\tan\beta$ from an accurate
measurement of the slope of the cross section in a region close to and
below 1 TeV would be possible for large values ($\gtrsim20$) of
$\tan\beta$, with an error that takes realistically into account the
"small" deviations of the next-to subleading term from a constant
value.\\

3) The next-to subleading term "essentially" depends, once given
$\tan\beta$ from the measured slope, only on the two mass parameters
$M_A$, $M_S$. Depending on the measured values of the parameters,
this could provide a simple but rather stringent test of the MSSM.\\

4) For the purposes of a test of the MSSM, the charged Higgs
production process exhibits special simplicity features that make it,
in our opinion, a very promising candidate. 
This property would suggest to extend our study to cases of non 
minimal SUSY models, in particular models whose Higgs couplings and 
structure are either different or richer.
We have completed our
examination using the dedicated code SESAMO; our next step will be
that of generalizing our analysis to the similar case of neutral Higgs
production. This will require an extension of our code, on which
work is already in progress.

\newpage

\appendix

\section{List of contributions and asymptotic expressions.}

In this Appendix we give the list of the various one loop diagrams 
that have to be retained for a practical computation (for example we
discard all diagrams that contribute proportionally to the light
lepton and quark masses). We follow the decomposition given in
eq.(\ref{A}).\\

\underline{Gauge boson self-energies}\\

These are the standard and supersymmetric bubbles and seagull
diagrams involving gauge bosons ($\gamma,~Z,~W$), goldstones,
ghosts, Higgses, fermions, charginos, neutralinos and sfermions.
They contribute the quantities $A^{c.t.}(e^+e^-\to H^+H^-)$
and $A^{s.e.}(e^+e^-\to H^+H^-)$, as explained in
refs.\cite{Hollik,HHHol}.\\

\underline{Initial vertices}\\

\begin{figure}
\centering
\epsfig{file=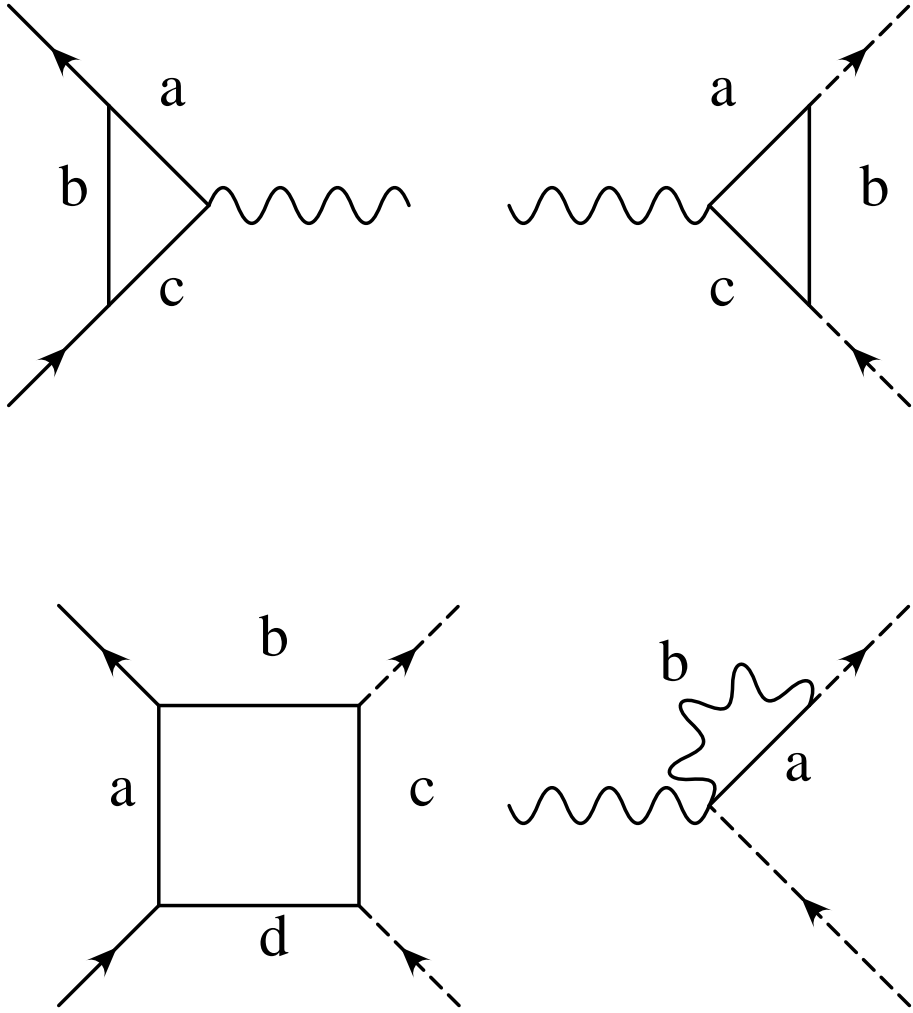,width=12cm}
\vskip 1cm\caption{Four classes of Feynman diagrams contributing $e^+e^-\to H^+H^-$ at one-loop. They are 
initial and final vertices (in the first row), boxes and final seagull diagrams (in the second row).}
\label{diagrams}
\end{figure}

The diagrams contributing $A^{in}(e^+e^-\to H^+H^-)$
are vertices with three internal lines sketched in Fig.~(\ref{diagrams})
and external $e^{\pm}$ self-energies. The list of vertices $(a,b,c)$ is:
($e,~\gamma,~e$),
($e,~Z,~e$), ($\nu,~W,~\nu$), ($W,~\nu,~W$), 
($\tilde{e},~\chi^0,~\tilde{e}$), 
($\tilde{\nu_{e}},~\chi^+,~\tilde{\nu_{e}}$),
($\chi^0,~\tilde{e},~\chi^0$), ($\chi^+,~\tilde{\nu_e},~\chi^+$).\\

\underline{final vertices}\\

The diagrams contributing $A^{fin}(e^+e^-\to H^+H^-$)
are vertices sketched in Fig.~(\ref{diagrams}) and external $H^\pm$
bubbles as well as seagull diagrams  involving the 
gauge boson-gauge boson-scalar-scalar couplings. The list of vertices
($a,b,c$) is: ($H,~\gamma,~H$), ($H,~Z,~H$), ($H,~W,~H$), ($W,~H,~W$),
($f,~f',~f$), ($\chi,~\chi,~\chi$),
($H,~H,~Z$), ($H,~H,~H$), ($\tilde{f},~\tilde{f'},~\tilde{f}$),
where $H$ and $\chi$ represent either charged or neutral states.
The list of seagull diagrams $(a,b)$ is 
$(H^+\gamma)$, $(H^+Z)$, $(H^0W)$, $(h^0W)$, $(A^0W)$.\\

\underline{boxes}\\

The contributions to $A^{box}(e^+e^-\to H^+H^-)$ are
box diagrams denoted clockwise by starting 
from the line running between $e^-$ and $e^+$ according to 
Fig.~(\ref{diagrams}). The list of boxes $(a,b,c,d)$ is: 
($\nu W H^0W$), ($e \gamma H^+\gamma$),~($eZH^+Z$),~($eZH^+\gamma$),~
($e\gamma H^+Z$),
($\tilde\nu \chi^+\chi^0\chi^+$),
($\tilde{e} \chi^0\chi^+\chi^0$),
($\chi^0\tilde{e_L}\tilde\nu\tilde{e_L}$),
($\chi^-\tilde\nu\tilde{e_L}\tilde\nu$),
($\tilde\nu\chi^+\chi^0\tilde{e_L}$),
($\tilde{e_L}\chi^0\chi^+\tilde\nu$).\\

{\bf Asymptotic expressions}\\

The complete expressions have been included in the code SESAMO.
Below we only give the results involving leading (DL)
and subleading (SL) logarithms.\\

Using the normalizations defined in Eq.(\ref{aLR}), we can write: 
\bq
a_\lambda = a_\lambda^{\rm Born} + \frac\alpha\pi(
\delta a_\lambda^{\rm s.e.}+\delta a_\lambda^{\rm in}
+\delta a_\lambda^{\rm fin})
\eq
where $\delta a_\lambda^{\rm s.e.}$, $\delta a_\lambda^{\rm in}$ and $\delta a_\lambda^{\rm fin}$
are the one-loop corrections to $a_\lambda^{\rm Born}$. \\

The asymptotic contributions from the intermediate 
$\gamma,~ Z$  self-energies are:

\bq
\delta a^{\rm s.e.}_L\to{1-2s^2_W+12s^4_W\over16s^4_Wc^4_W}
\ \log q^2~~~~~~~~~
\delta a^{\rm s.e.}_R\to{11\over8c^4_W}
\ \log q^2
\eq

those from initial $e^+e^-$ lines:

\bqa
&&\delta a^{\rm in}_L\to{1\over64s^4_Wc^4_W}\ 
\left(2\log q^2-\log^2{q^2\over M^2_Z}\right)
+{1\over32s^4_Wc^2_W}\
\left(2\log q^2-\log^2{q^2\over M^2_W}\right)
\nonumber\\
&&
\delta a^{\rm in}_R\to{1\over8c^4_W}\ 
\left(2\log q^2-\log^2{q^2\over M^2_Z}\right)
\eqa

and those from final $H^+H^-$ lines and boxes:

\bqa
&&\delta a^{\rm fin}_L\to{1\over64s^4_Wc^4_W}
\ \left(2\log q^2-\log^2{q^2\over M^2_Z}\right)
+{1\over32s^4_Wc^2_W}\ 
\left(2\log q^2-\log^2{q^2\over M^2_W}\right)\nonumber\\
&&-~\frac{3}{32 s_w^4c^2_W M_W^2}\ 
(m_t^2\cot^2\beta+m_b^2\tan^2\beta)\ \log q^2 \nonumber\\
&&-~{1\over16s^4_Wc^4_W}\  \log {1-cos\vartheta\over 1+cos\vartheta} \cdot \log q^2
-~{1\over4s^4_W}\ \log {1-cos\vartheta\over2}\cdot \log q^2
\eqa
\bqa
&&
\delta a^{\rm fin}_R\to{1\over8c^4_W}
\left(2\log q^2-\log^2{q^2\over M^2_Z}\right)
+{1\over16s^2_Wc^2_W}
\left(2\log q^2-\log^2{q^2\over M^2_W}\right)\nonumber\\
&&-~\frac{3}{16 s_w^2c^2_W M_W^2}\
(m_t^2\cot^2\beta+m_b^2\tan^2\beta)\ \log q^2\nonumber\\
&&
-~{1\over4c^4_W}\ \log {1-cos\vartheta\over 1+cos\vartheta}\cdot\log q^2
\eqa
\noindent
where $\vartheta$ is the c.m. angle between initial $e^-$
and final $H^-$ momenta.

\newpage

\newpage

\newpage

\newpage

\newpage

\newpage

\newpage

\newpage

\newpage

\newpage

\newpage

\newpage

\end{document}